\documentclass[a4paper,12pt]{article}
\usepackage{amsmath,amssymb}
\usepackage[english]{babel}

\title{\vspace{-2cm}{\small\sl
\rightline{Submitted to Funct. Anal. Appl.}}\bigskip
On the relation between multifield and multidimensional integrable
equations}
\author{V.E. Adler}
\date{Ufa Institute of Mathematics, 112 Chernyshevsky str., \\
Ufa 450077, Russia \\ e-mail: adler@imat.rb.ru}

\def\Real{\mathbb{R}}

\def\mat{\mathop{\rm Mat}\nolimits}
\def\<{\langle}
\def\>{\rangle}
\let\old=\phi \let\phi=\varphi \let\varphi=\old
\newtheorem{theorem}{Theorem}

\begin{document} \maketitle \thispagestyle{empty}

\begin{abstract}
The new examples are found of the constraints which link the
1+2-dimensional and multifield integrable equations and lattices. The
vector and matrix generalizations of the Nonlinear Schr\"odinger equation
and the Ablowitz-Ladik lattice are considered among the other multifield
models. It is demonstrated that using of the symmetries belonging to the
hierarchies of these equations leads, in particular, to the KP equation
and twodimensional analogs of the dressing chain, Toda lattice and
dispersive long waves equations. In these examples the multifield equation
and its symmetry have meaning of the Lax pair for the corresponding
twodimensional equation under some compatible constraint between field
variables and eigenfunctions.
\end{abstract}

%--------------------------------------------------------------------------
\section{Introduction. Vector NLS equation}\label{vec}

Let us consider the vector generalization of the Nonlinear Schr\"odinger
equation (NLS) \cite{Manakov,Fordy-Kulish} and its third order symmetry:
\begin{gather}
\label{vec:y}
   \psi_y= \psi_{xx} +2\<\psi,\phi\>\psi, \quad
  -\phi_y= \phi_{xx} +2\<\psi,\phi\>\phi, \\
\label{vec:t}
 \begin{split}
   \psi_t&= \psi_{xxx} +3\<\psi,\phi\>\psi_x +3\<\psi_x,\phi\>\psi, \\
   \phi_t&= \phi_{xxx} +3\<\psi,\phi\>\phi_x +3\<\psi,\phi_x\>\phi
 \end{split}
\end{gather}
where $\psi,\phi\in\Real^N$ and the brackets $\<\,,\>$ denote scalar
product. It is an easy exercise to check that the quantities
\begin{equation}\label{vec:uq}
  u=-2\<\psi,\phi\>,\quad q=2\<\psi,\phi_x\>-2\<\psi_x,\phi\>
\end{equation}
solve, in virtue of these equations and independently on the dimension $N,$
the Kadomtsev-Petviashvili equation (KP)
\begin{equation}\label{vec:KP}
  4u_t=u_{xxx}-6uu_x+3q_y, \quad q_x=u_y.
\end{equation}
This observation was done in the papers \cite{Konopelchenko-Strampp91,KSS}
for the first time. It reduces the problem of solving of the
1+2-dimensional equation to the simpler problem of construction of the
common solution of the pair of 1+1-dimensional systems, although of the
large size. Apparently, this trick is a sort of nonlinear analog of the
separation of variables method. It is rather effective and even the
simlest case $N=1$ have brought to some new exact solutions of KP
\cite{Cheng-Li91}.

We mention also the further contributions \cite{Cheng-Li92, Xu-Li,
Konopelchenko-Strampp92a, Konopelchenko-Strampp92b}.  Within this approach
eqs.~(\ref{vec:y}), (\ref{vec:t}) are treated as the pair of the formally
conjugated auxiliary linear problems for the KP equation
\begin{gather}
\label{vec:auxy}
  \psi_y=\psi_{xx}-u\psi, \quad   -\phi_y = \phi_{xx}-u\phi, \\
\label{vec:auxt}
  \psi_t=\psi_{xxx}-\frac32u\psi_x-\frac34(u_x+q)\psi, \quad
  \phi_t=\phi_{xxx}-\frac32u\phi_x-\frac34(u_x-q)\phi
\end{gather}
and the relations (\ref{vec:uq}) are treated as the constraint between the
potential and the eigenfunctions, which is consistent with the dynamics on
$y$ and $t$. Bearing this interpretation in mind we always denote the
multifield variables (vectors or matrices) as $\psi,\phi.$

If the 1+2-dimensional equation is given then the finding of the
corresponding 1+1-dimensional systems can be rather nontrivial task, even
if the $L-A$-pair is known.  Several solutions can exist. For example, it
turns out that eqs.~(\ref{vec:KP}) -- (\ref{vec:auxt}) admit also other
compatible constraint:
\begin{gather*}
 u= 2\<\psi_x,\phi_x\>
    -2\frac{\<\psi,\phi_x\>\<\psi_x,\phi\>}{\<\psi,\phi\>+1}, \\
 q= 2\<\psi_{xx},\phi_x\> -2\<\psi_x,\phi_{xx}\>
    +2\frac{ \<\psi,\phi_x\>_x\<\psi_x,\phi\>
            -\<\psi,\phi_x\>\<\psi_x,\phi\>_x }
           {\<\psi,\phi\>+1}                                  \\
    +2\frac{ \<\psi,\phi_x\>\<\psi_x,\phi\>
             (\<\psi,\phi_x\>-\<\psi_x,\phi\>) }
           {(\<\psi,\phi\>+1)^2}
\end{gather*}
(its origin will be explained in Section \ref{AL}). One more
representation for the KP equation is given in the Example 3 below, but in
this case the equations for $\psi$ and $\phi$ are not conjugated.

On the other hand, now we know a large number of separate examples and the
whole classes of integrable multifield equations (see, e.g. the papers
\cite{Fordy-Kulish, Fordy, Svinolupov89, Svinolupov92, Svinolupov93,
Svinolupov-Sokolov, Olver-Sokolov1, Olver-Sokolov2} and the review
\cite{HSY}) and the natural question arize, which 1+2-dimensional
equations are related to them. One of the goals of this paper is to obtain
some new examples of such relation.

Our second goal is to apply this approach to the differential-difference
equations (lattices). For example, let us consider the lattice generated
by iterations of the B\"acklund transformation (BT) for the equation
(\ref{vec:y}):
\begin{equation}\label{vec:NLSBT}
  \psi_x=\psi_1+\beta\psi+\<\psi,\phi_1\>\psi,\quad
 -\phi_x=\phi_{-1}+\beta_{-1}\phi+\<\psi_{-1},\phi\>\phi,
\end{equation}
where $\beta_j$ are arbitrary scalar parameters \cite{SY91,Ad94}. Here and
further on, when working with the lattices, we omit for short the dummy
integer subscript $j,$ so that $\psi$ means $\psi_j,$ $\psi_1$ means
$\psi_{j+1}$ and so on.

One obtains, after introducing the quantities
\[
  f=-\<\psi,\phi_1\>-\beta, \quad
  p=\<\psi,\phi_{1,x}\>-\<\psi_x,\phi_1\>+\<\psi,\phi_1\>^2-\beta^2,
\]
the well known Miura-type transformations \cite{Konopelchenko82,Jimbo-Miwa}
\[
  u_1=u+2f_x,\quad u=f^2-f_x+p,\quad p_x=f_y
\]
which connect eq. (\ref{vec:KP}) with the modified Kadomtsev-Petviashvili
equation (mKP)
\begin{equation}\label{vec:mKP}
  4f_t=f_{xxx}-6(f^2+p)f_x+3p_y,\quad p_x=f_y.
\end{equation}
The variables $f,p$ satisfy the lattice
\begin{equation}\label{vec:mKPBT}
 f_{1,x}+f_x=f^2_1-f^2+p_1-p,\quad p_x=f_y.
\end{equation}
Investigation in this direction can be useful for constructing of the
explicit solutions of 1+2-dimensional equations (however, it falls beyond
the scope of the prersent paper). Let us recall that if the dependence on
$y$ is neglected then eq. (\ref{vec:KP}) turns into the Korteweg-de Vries
equation (KdV), and the lattice (\ref{vec:mKPBT}) turns into the dressing
chain. It was proved in \cite{Veselov-Shabat} that the dressing chain
under the periodicity condition $f_j=f_{j+K}$ is a completely integrable
finite-dimensional Hamiltonian system, and all the $k$-gap solutions of KdV
can be derived from it at $K=2k+1.$  In the two-dimensional case, some
large class of the KP solutions can be obtained under the periodic
condition $\psi_j=\psi_{j+K},$ $\phi_j=\phi_{j+K}$ for the lattice
(\ref{vec:NLSBT}) which brings to the Hamiltonian system with $KN$ degrees
of freedom.  The question if all algebraic-geometric solutions of KP can
be obtained in such a way remains open. The multisoliton solutions can be
constructed also by means of the nonlinear superposition principle for the
lattice (\ref{vec:NLSBT}) \cite{Ad94}.

The twodimensional examples considered in this paper include several KP-
and mKP-like equations, dispersive long waves equations and, in the
discrete case, the twodimensional dressing chain and Toda lattice.
Moreover, the matrix versions of these equations are considered.  On the
other hand we have the vector and matrix generalizations of NLS, modified
NLS and derivative NLS. Probably, most beautiful example is related to the
vector generalization of the Ablowitz-Ladik lattice which was introduced
recently in the papers \cite{AOT}.  It should be mentioned that the scalar
Ablowitz-Ladik lattice has been already used for constructing of the
solutions of 1+2-dimensional equation \cite{Vekslerchik96}.  As a rule, the
order of nonlinear terms in multifield equations is odd, however exceptions
are possible (see the last Section).

%--------------------------------------------------------------------------
\section{Jordan NLS}\label{NLS}

We start from the further generalizations of NLS which were proved to be
integrable in \cite{Svinolupov92}. These system are of the form
\begin{equation}\label{NLS:y}
  \psi_y= \psi_{xx}+2\{\psi\phi\psi\}, \quad
 -\phi_y= \phi_{xx}+2\{\phi\psi\phi\}
\end{equation}
where the braces denote triple product which satisfies the axioms of the
Jordan pair:
\begin{equation}\label{NLS:J}
\begin{split}
 \{abc\}&=\{cba\}, \\
 \{ab\{cde\}\}-\{cd\{abe\}\}&=\{\{abc\}de\}-\{c\{bad\}e\}.
\end{split}
\end{equation}
Below we consider only the vector and matrix examples which correspond to
the following, most important, rules of multiplication:
\begin{align}
  2\{abc\}&= \<a,b\>c +\<c,b\>a, \qquad
    a,b,c\in\Real^N,                        \label{NLS:mult1}\\
   \{abc\}&= \<a,b\>c +\<c,b\>a -\<a,c\>b, \qquad
    a,b,c\in\Real^N,                        \label{NLS:mult2}\\
  2\{abc\}&= abc + cba,\qquad
    a,c\in\mat_{M,N},\quad b\in\mat_{N,M}.  \label{NLS:mult3}
\end{align}
Nevertheless, it is convenient to formulate some statements in general
algebraic terms. In particular, the eq.~(\ref{NLS:y}) admits the third
order symmetry
\begin{equation}\label{NLS:t}
 \psi_t= \psi_{xxx}+6\{\psi\phi\psi_x\}, \quad
 \phi_t= \phi_{xxx}+6\{\phi\psi\phi_x\}
\end{equation}
and the B\"acklund transformation \cite{SY91,Ad94}
\begin{equation}\label{NLS:BT}
  \psi_x= \psi_1+\beta\psi+\{\psi\phi_1\psi\}, \quad
 -\phi_x= \phi_{-1}+\beta_{-1}\phi+\{\phi\psi_{-1}\phi\}.
\end{equation}

%--------------------
\paragraph{Example 1.} Obviously, the example discussed in Introduction
corresponds to the multiplication (\ref{NLS:mult1}) which is the
particular case of the matrix multiplication (\ref{NLS:mult3}) at $M=1$.
In the general case consider the matrix NLS
\[
  \psi_y=\psi_{xx}+2\psi\phi\psi, \quad -\phi_y=\phi_{xx}+2\phi\psi\phi
\]
where $\psi\in\mat_{M,N},$ $\phi\in\mat_{N,M}.$ Let us introduce the
$M\times M$ matrices
\begin{gather*}
 u=-2\psi\phi, \quad q=2\psi\phi_x-2\psi_x\phi, \\
 f=-\psi\phi_1-\beta I_M, \quad
 p=\psi\phi_{1,x}-\psi_x\phi_1+\psi\phi_1\psi\phi_1-\beta^2I_M
\end{gather*}
where $I_M$ is the identity matrix, then using the symmetry (\ref{NLS:t})
yields the matrix KP equation (here and below the square brackets denote
commutator $[a,b]=ab-ba$)
\begin{equation}\label{NLS:matKP}
 4u_t= u_{xxx}-3(uu_x+u_xu-q_y+[q,u]), \quad  q_x=u_y
\end{equation}
while the lattice (\ref{NLS:BT}) yields the B\"acklund transformation
\begin{gather*}
  u_1=u+2f_x,\quad u=f^2-f_x+p,\\
  f_{1,x}+f_x=f^2_1-f^2+p_1-p,\quad p_x=f_y+[p,f].
\end{gather*}
It should be stressed that, in contrast to the scalar case, the matrix $f$
does not satisfy the matrix mKP equation.

%--------------------
\paragraph{Example 2.} The version of the vector NLS equation related to
the multiplication (\ref{NLS:mult2}) was introduced in
\cite{Kulish-Sklyanin} for the first time:
\[
  \psi_y=\psi_{xx}+2\<\psi,\phi\>\psi-\<\psi,\psi\>\phi, \quad
 -\phi_y=\phi_{xx}+2\<\psi,\phi\>\phi-\<\phi,\phi\>\psi.
\]
In this example the quantities
\[
 u=-4\<\psi,\phi\>,\quad v=2\<\psi,\psi\>,\quad
 w= 2\<\phi,\phi\>,\quad q=4\<\psi,\phi_x\>-4\<\psi_x,\phi\>
\]
satisfy, in virtue of (\ref{NLS:t}), the 1+2-dimensional system
\begin{align*}
  4u_t&= u_{xxx}-6uu_x+3q_y+6(vw)_x, \quad q_x=u_y, \\
 -2v_t&= v_{xxx}-3v_{xy}-3uv_x+3vq, \\
 -2w_t&= w_{xxx}+3w_{xy}-3uw_x-3wq
\end{align*}
which generalize KP (reduction $vw=0$).
\smallskip

In conclusion we notice that actually the equation (\ref{NLS:y}) admits
the infinite hierarchy of the higher symmetry (it can be constructed, for
example, by means of the recursion operator \cite{Svinolupov92}) and using
them instead of (\ref{NLS:t}) brings to the higher symmetries of the
corresponding KP-like equations. In some sense this solves the problem of
nonlocalities which is a serious technical obstacle in the theory of
twodimensional equations. However, we always consider only the simplest
representatives of the integrable hierarchies in the further examples.

%--------------------------------------------------------------------------
\section{Derivative NLS}\label{DNLS}

There are two different scalar systems known under the name of the
derivative NLS equation:
\[
 \left\{\begin{array}{r}
   \psi_y= \psi_{xx} +2(\psi^2\phi)_x,\\
  -\phi_y= \phi_{xx} -2(\psi\phi^2)_x
 \end{array}\right.
 \qquad \mbox{and} \qquad
 \left\{\begin{array}{r}
   \psi_y= \psi_{xx}+2\psi\phi\psi_x, \\
  -\psi_y= \phi_{xx}-2\psi\phi\phi_x
 \end{array}\right.
\]
which were introduced in the papers \cite{Kaup-Newell} and \cite{CLL}
correspondingly.  They are related by differential substitution.  In the
multifield case the difference becomes more deep: the first system admits
the natural generalization in the Jordan pairs, as NLS, while the second
one is generalized in associative algebras. We consider the second version
first.

\begin{theorem}[Olver, Sokolov\cite{Olver-Sokolov1}]\label{prop:DNLS1}
Let $\psi,\phi$ belong to an associative algebra, then the system
\begin{equation}\label{DNLS1:y}
  \psi_y= \psi_{xx} +2\psi_x\phi\psi, \quad
  \phi_y=-\phi_{xx} +2\phi\psi\phi_x
\end{equation}
possesses the third order symmetry
\begin{equation}\label{DNLS1:t}
\begin{split}
 \psi_t&= \psi_{xxx} +3\psi_{xx}\phi\psi +3\psi_x\phi\psi_x
                     +3\psi_x\phi\psi\phi\psi, \\
 \phi_t&= \phi_{xxx} -3\phi\psi\phi_{xx} -3\phi_x\psi\phi_x
                     +3\phi\psi\phi\psi\phi_x.
\end{split}
\end{equation}
\end{theorem}

We need some extension of this result. Namely, it is easy to see that
actually the associativity of multiplication is sufficient, without the
requirement that $\psi,\phi$ belong to the same space.  In particular, we
consider the case $\psi\in\mat_{M,N},$ $\phi\in\mat_{N,M}.$  Apparently,
the equations (\ref{DNLS1:y}), (\ref{DNLS1:t}) can be treated as an
auxiliary linear problem in two ways, one leads to the matrix KP and
another one to matrix mKP equation.

%--------------------
\paragraph{Example 3.} Let us consider $M\times M$ matrices
\[
  u= -2\psi_x\phi, \quad
  q= 2\psi_x\phi_x -2\psi_{xx}\phi - 4\psi_x\phi\psi\phi
\]
then $\psi$ satisfy equations of the form
\[
  \psi_y= \psi_{xx} -u\psi, \quad
  \psi_t=\psi_{xxx}-\frac32u\psi_x-\frac34(u_x+q)\psi
\]
and, in virtue of their consistency, $u$ and $q$ satisfy
eq.~(\ref{NLS:matKP}).  In particular, a new representation of the scalar
KP arise at $M=1.$

%--------------------
\paragraph{Example 4.} Now let us consider the $N\times N$ matrices
\[
  f=\phi\psi, \quad p=\phi\psi_x - \phi_x\psi + f^2.
\]
They satisfy the matrix mKP equation
\begin{equation}\label{DNLS1:matmKP}
 4f_t= f_{xxx} +3([f_{xx},f] -2ff_xf +p_y +[p,f^2] +f_xp+pf_x), \quad
 f_y= p_x +[p,f]
\end{equation}
in virtue of (\ref{DNLS1:y}), (\ref{DNLS1:t})

Next, let us discuss the multifield analogs of the Kaup-Newell system.
They were studied, e.g. in \cite{Fordy,ASY}.

\begin{theorem}
Let triple product $\{\ \}$ satisfy axioms (\ref{NLS:J}) then the system
\begin{equation}\label{DNLS2:y}
  \psi_y= \psi_{xx}+2\{\psi\phi\psi\}_x, \quad
  \phi_y=-\phi_{xx}+2\{\phi\psi\phi\}_x
\end{equation}
admits the symmetry of third order
\begin{equation}\label{DNLS2:t}
\begin{split}
 \psi_t&= (\psi_{xx} +6\{\psi\phi\psi_x\}
             +6\{\psi\{\phi\psi\phi\}\psi\})_x, \\
 \phi_t&= (\phi_{xx} -6\{\phi\psi\phi_x\}
             +6\{\phi\{\psi\phi\psi\}\phi\})_x.
\end{split}
\end{equation}
\end{theorem}

This statement was proved in \cite{ASY} in the case of the Jordan triple
systems, but, as before, it remains valid also for the Jordan pairs, that
is when $\psi$ and $\phi$ belong to the different spaces. We just write
down the resulting two-dimensional equations.

%--------------------
\paragraph{Example 5.} In the case of multiplication (\ref{NLS:mult3}) the
matrices
\[
  f=\phi\psi, \quad p=\phi\psi_x - \phi_x\psi + 3f^2
\]
solve, in virtue of (\ref{DNLS2:y}), (\ref{DNLS2:t}) the same equation
(\ref{DNLS1:matmKP}). This coincidence, although unexpected at a first
glance, is easily explained if one consider the corresponding auxiliary
linear problems, which are formally conjugated.

%--------------------
\paragraph{Example 6.} In the case of multiplication (\ref{NLS:mult2}) the
quantities
\[
  f=-2\<\psi,\phi\>,\ g=\<\psi,\psi\>,\ h=\<\phi,\phi\>,\
  p=2\<\psi,\phi_x\>-2\<\psi_x,\phi\> -3f^2 +6gh
\]
satisfy the system of equations
\begin{align*}
  4f_t&= f_{xxx} +3(p-2gh)_y +6(g_{xx}h-gh_{xx}) \\
      &\qquad\qquad -2(6fgh+f^3)_x -6pf_x,  \quad p_x= f_y   \\
 -2g_t&= g_{xxx} -3g_{xy} +6f(g_y-g_{xx}) +3(g(p-f_x))_x \\
      &\qquad\qquad +6g^2h_x +6gff_x +9g_xf^2, \\
 -2h_t&= h_{xxx} +3h_{xy} +6f(h_y+h_{xx}) +3(h(p+f_x))_x \\
      &\qquad\qquad +6h^2g_x +6hff_x +9h_xf^2
\end{align*}
which is reduced to mKP (\ref{vec:mKP}) at $gh=0.$

%--------------------------------------------------------------------------
\section{Modified Jordan NLS}\label{mNLS}

Let us consider the lattice (\ref{NLS:BT}), assuming $\beta_j=0$ for sake
of simplicity, and make renaming $\psi=\tilde\psi,$ $\phi_1=\tilde\phi.$
This give rise to the differential substitution
\[
  \psi=\tilde\psi,\quad
  \phi=-\tilde\phi_x-\{\tilde\phi\tilde\psi\tilde\phi\}
\]
and one can easily see that eq. (\ref{NLS:y}) can be rewritten in new
variables. The result is the modified Jordan NLS (the tilde is omitted)
\cite{Ablowitz-Segur}
\begin{equation}\label{mNLS:y}
\begin{split}
  \psi_y&= \psi_{xx}-2\{\psi\phi_x\psi\}-2\{\psi\{\phi\psi\phi\}\psi\},\\
 -\phi_y&= \phi_{xx}+2\{\phi\psi_x\phi\}-2\{\phi\{\psi\phi\psi\}\phi\}
\end{split}
\end{equation}
and the chain of its B\"acklund transformations
\begin{equation}\label{mNLS:BT}
 \psi_x=\psi_1+\{\psi\phi\psi\}, \quad -\phi_x=\phi_{-1}+\{\phi\psi\phi\}.
\end{equation}
We will not rewrite the symmetry (\ref{NLS:t}) since it is clear without
calculations that the resulting twodimensional equations are exactly the
same as in the Section \ref{NLS} and the difference is only in the
formulae which define the constraints between potential and eigenfunctions.

The new examples arise if one consider the time with a ``negative number''
from the hierarchy of eq.~(\ref{mNLS:y}) symmetries.

\begin{theorem}
For any Jordan pair equations (\ref{mNLS:y}) define the symmetry of the
following hyperbolic system
\begin{equation}\label{mNLS:xz}
 \psi_{xz}= 2\{\psi\phi\psi_z\}-\psi, \quad
 \phi_{xz}=-2\{\phi\psi\phi_z\}-\phi.
\end{equation}
\end{theorem}
The proof is straightforward and use only identities (\ref{NLS:J}).

%--------------------
\paragraph{Example 7.} If multiplication is defined by formula
(\ref{NLS:mult1}) then eqs.~(\ref{mNLS:y}), (\ref{mNLS:xz}) take the form
\begin{gather}
\label{mNLS:y1}
 \begin{split}
  \psi_y&= \psi_{xx} -2(\<\psi,\phi_x\>+\<\psi,\phi\>^2)\psi, \\
 -\phi_y&= \phi_{xx} +2(\<\psi_x,\phi\>-\<\psi,\phi\>^2)\phi,
 \end{split} \\
\label{mNLS:xz1}
  \psi_{xz}= \<\psi,\phi\>\psi_z +(\<\psi_z,\phi\>-1)\psi, \quad
  \phi_{xz}=-\<\psi,\phi\>\phi_z -(\<\psi,\phi_z\>+1)\phi.
\end{gather}
In virtue of them the quantities
\[
  u=\<\psi,\phi\>,\quad v=\<\psi_z,\phi\>-1,\quad q=\<\psi,\phi_x\>+u^2
\]
satisfy the well known two-dimensional generalization of the dispersive
long waves equations (DLW) \cite{BLP1, BLP2, Konopelchenko88}
\begin{equation}\label{mNLS:DLW}
  u_y=(u_x+u^2-2q)_x, \quad -v_t=(v_x-2uv)_x,\quad q_z=v_x.
\end{equation}
The lattice (\ref{mNLS:BT}) takes form
\[
  \psi_x=\psi_1+\<\psi,\phi\>\psi,\quad
 -\phi_x=\phi_{-1}+\<\psi,\phi\>\phi
\]
and is the member of the hierarchy which was studied in details in the
paper \cite{MRZ}.  In particular, it was proved there that this lattice
admits the symmetry of the form
\begin{equation}\label{mNLS:z}
  \psi_z=\frac{\psi_{-1}}{\<\psi_{-1},\phi\>-1},\quad
 -\phi_z=\frac{\phi_1}{\<\psi,\phi_1\>-1}.
\end{equation}
One can easily check that this pair of lattices define the BT for the
system (\ref{mNLS:xz1}). Moreover, the variables $u,v$ satisfy the
two-dimensional Toda lattice
\begin{equation}\label{mNLS:Toda}
   u_z=v-v_1,\quad v_x=v(u-u_{-1}),
\end{equation}
which define the BT for the system (\ref{mNLS:DLW})
\cite{Konopelchenko91,SY97}.

%--------------------
\paragraph{Example 8.} Analogs of the lattice (\ref{mNLS:z}) exist for
other Jordan pairs as well. Let $\psi\in\mat_{M,N},$ $\phi\in\mat_{N,M},$
then the following differentiations commute in the matrix case
(\ref{NLS:mult3}):
\begin{gather*}
  \psi_x= \psi_1+\psi\phi\psi,               \quad
 -\phi_x= \phi_{-1}+\phi\psi\phi,            \\
  \psi_z= (\psi_{-1}\phi-I_M)^{-1}\psi_{-1}, \quad
 -\phi_z= \phi_1(\psi\phi_1-I_M)^{-1},       \\
  \psi_y= \psi_{xx}-2\psi\phi_x\psi-2\psi\phi\psi\phi\psi,  \quad
 -\phi_y= \phi_{xx}+2\phi\psi_x\phi-2\phi\psi\phi\psi\phi
\end{gather*}
and the matrices of the $M\times M$ size
\[
  u=\psi\phi,\quad
  v=\psi_z\phi-I_M=(\psi_{-1}\phi-I_M)^{-1},\quad
  q=\psi\phi_x+u^2=-\psi\phi_{-1}
\]
satisfy the nonabelian twodimensional DLW equations and Toda lattice
\cite{Konopelchenko91}
\begin{gather*}
  u_y=u_{xx}+2(u_xu-q_x+[u,q]), \quad
 -v_y=v_{xx}-2(uv)_x-2[v,q], \quad q_z=v_x, \\
  u_z=v-v_1,\quad v_x=uv-vu_{-1}.
\end{gather*}

%--------------------------------------------------------------------------
\section{Vector Ablowitz-Ladik lattice}\label{AL}

Probably one of the most beautiful examples is related to the multifield
Ablowitz-Ladik lattices. It yields the same twodimensional equations as in
the previous section, but is more symmetric.

\begin{theorem}
Let $\psi\in\mat_{M,N},$ $\phi\in\mat_{N,M}$ then the lattices
\begin{equation}\label{AL:xz}
 \left\{\begin{array}{rl}
   \psi_x=& \psi_{-1} + \psi_{-1}\phi\psi, \\
  -\phi_x=& \phi_1 + \phi\psi\phi_1,
 \end{array}\right.
 \qquad
 \left\{\begin{array}{rl}
   \psi_z=& \psi_1 + \psi\phi\psi_1, \\
  -\phi_z=& \phi_{-1} + \phi_{-1}\psi\phi
 \end{array}\right.
\end{equation}
commute and the variables $\psi,\phi$ satisfy the following system in
virtue of these lattices:
\begin{equation}\label{AL:PLR}
\begin{split}
  \psi_{xz}&= \psi_x\phi(\psi\phi+I_M)^{-1}\psi_z +\psi\phi\psi +\psi, \\
  \phi_{xz}&= \phi_z(\psi\phi+I_M)^{-1}\psi\phi_x +\phi\psi\phi +\phi.
\end{split}
\end{equation}
\end{theorem}

In the scalar case the sum of the flows (\ref{AL:xz}) and the dilation
symmetry $\psi_\tau=-2\psi,$ $\phi_\tau=2\phi$ is the Ablowitz-Ladik
lattice \cite{AL} and (\ref{AL:PLR}) is Pohlmeyer-Lund-Regge system (this
observation due to \cite{SY90}). The second order symmetries of the
eq.~(\ref{AL:xz}) hierarchy is nothing but derivative NLS equations
(cf~(\ref{DNLS1:y})):
\begin{gather}
 \psi_y =  \psi_{xx} + 2\psi_x\phi_1\psi,       \quad
 \phi_y = -\phi_{xx} + 2\phi\psi_{-1}\phi_x,    \label{AL:y}\\
 \psi_\eta =  \psi_{zz} + 2\psi\phi_{-1}\psi_z, \quad
 \phi_\eta = -\phi_{zz} + 2\phi_z\psi_1\phi.    \label{AL:eta}
\end{gather}
Of course, one should eliminate the variables $\psi_{\pm1},$ $\phi_{\pm1}$
in virtue of the lattices in order two treat these equations as higher
symmetries of the system (\ref{AL:PLR}).

I will consider only the vector case $M=1:$
\begin{gather}
\label{AL:vec}
 \left\{\begin{array}{rl}
   \psi_x=& \psi_{-1} + \<\psi_{-1},\phi\>\psi, \\
  -\phi_x=& \phi_1 + \<\psi,\phi_1\>\phi,
 \end{array}\right.
  \quad
 \left\{\begin{array}{rl}
   \psi_z=& \psi_1 + \<\psi,\phi\>\psi_1,   \\
  -\phi_z=& \phi_{-1} + \<\psi,\phi\>\phi_{-1},
 \end{array}\right.
 \\
\label{AL:PLRvec}
 \left\{\begin{array}{rl}
   \psi_{xz}=& \dfrac{\<\psi_x,\phi\>}{\<\psi,\phi\>+1}\psi_z
                 +\<\psi,\phi\>\psi +\psi,  \\
   \phi_{xz}=& \dfrac{\<\psi,\phi_x\>}{\<\psi,\phi\>+1}\phi_z
                 +\<\psi,\phi\>\phi +\phi.
 \end{array}\right.
\end{gather}
Now the sum of the lattices (\ref{AL:vec}) (plus dilation symmetry) defines
the so-called ``asymmetric discretization of vector NLS'' introduced in
\cite{AOT}.  It should be mentioned that another discretization was
considered in these papers as well, which can be obtained by summing the
second lattice (\ref{AL:vec}) and the lattice symmetric with respect to
reflection $j\to-j.$  Integrability of this version was established in
\cite{TUW}, however the question about the structure of the higher
symmetries remains open.

The quantities
\[
  v=\<\psi,\phi\>+1,\quad  u=-\<\psi,\phi_1\>=\<\psi,\phi_x\>/v
\]
satisfy the Toda lattice (\ref{mNLS:Toda}) in virtue of (\ref{AL:vec}). In
order to reproduce the DLW equations, it is sufficient to consider the
symmetries (\ref{AL:y}), (\ref{AL:eta}). First one takes the form
\begin{gather*}
 \psi_y=\psi_{xx}-2q\psi, \quad -\phi_y=\phi_{xx}-2q\phi, \quad
 q=\<\psi_x,\phi_x\>-\frac{\<\psi,\phi_x\>\<\psi_x,\phi\>}{\<\psi,\phi\>+1}
\end{gather*}
after elimination of $\psi_{-1},$ $\phi_1$ and brings to
eq.~(\ref{mNLS:DLW}), up to the change of the sign of $y.$  Notice, that
this formula coincide, up to the renaming of the potential, with the
auxiliary linear problem (\ref{vec:auxy}) for KP.  It can be easily proved
that the form of the second auxiliary problem (\ref{vec:auxt}) is uniquely
derived from the compatibility condition and yields the third order
symmetry of the system (\ref{AL:PLRvec}). This gives the new example of
the constraint for KP, which was displayed in Introduction.

The symmetry (\ref{AL:eta}) is rewritten as follows
\[
 \psi_\eta=  \psi_{zz}-2\frac{\<\psi,\phi_z\>}{\<\psi,\phi\>+1}\psi_z,\quad
 \phi_\eta= -\phi_{zz}+2\frac{\<\psi_z,\phi\>}{\<\psi,\phi\>+1}\phi_z
\]
and denoting $p=\<\psi_z,\phi\>/v$ one obtains another equation from the
DLW hierarchy:
\[
  u_\eta=u_{zz}-2v_z+2pu_z,\quad v_\eta=-(v_z-2pv)_z,\quad p_x=u_z.
\]

%--------------------------------------------------------------------------
\section{Concluding examples}

Notice, that the order of nonlinear terms was odd for all examples of
multifield equations considered above, while the constraints were defined
by expressions with even order products of $\psi$-functions. Here we
present two examples with the different structure of nonlinearities.

%--------------------
\paragraph{Example 9.} Let us consider the generalization of the KdV
equation which corresponds to the Jordan algebra $D_N$ \cite{Svinolupov93}:
\[
  u_y= u_{xxx}+3(u^2 -\<\psi,\psi\>)_x, \quad
  \psi_y= \psi_{xxx} +6(u\psi)_x, \quad u\in\Real,\ \psi\in\Real^N.
\]
It possesses the fifth order symmetry
\begin{align*}
 u_t&= u_{xxxxx} +5(2uu_{xx} +u^2_x +2u^3 -6u\<\psi,\psi\>
                    -2\<\psi,\psi_{xx}\> -\<\psi_x,\psi_x\>)_x, \\
\psi_t&= \psi_{xxxxx} +10(u\psi_{xx}+u_x\psi_x+u_{xx}\psi
                          +3u^2\psi -\<\psi,\psi\>\psi)_x
\end{align*}
and the quantities
\[
  v=6u, \quad q=6u_{xx}+18u^2-18\<\psi,\psi\>
\]
satisfy, in virtue of these equations, the twodimensional generalization of
Sawada-Kotera equation \cite{SK}
\[
 -9v_t= v_{xxxxx} +5(vv_{xxx}+v_xv_{xx}+v^2v_x-v_{xxy}-vv_y-qv_x-q_y),
 \quad q_x=v_y.
\]

%--------------------
\paragraph{Example 10.} The equation
\[
  \psi_y= \psi_{xx} +2\psi\psi_x +[\psi,\psi^2]
\]
on arbitrary left-symmetric algebra was proved to be integrable in
\cite{Svinolupov89}. It admits the infinite hierarchy of higher
symmetries, simplest one is of the form
\[
 \psi_t= \psi_{xxx} +3(\psi\psi_{xx}+\psi^2_x
        +[\psi,\psi\psi_x]+\psi(\psi_x\psi)) +[\psi,\psi\psi^2].
\]
This equation is a multifield generalization of Burgers equation and is
linearizable via Cole-Hopf type substitution.

An example of left-symmetric multiplication is given by formula
\[
 \psi\phi=\<\psi,c\>\phi+\<\psi,\phi\>c, \quad \psi,\phi,c\in\Real^n
\]
where vector $c$ is fixed. One can check that in this case the quantities
\[
  u=\<\psi,c\>,\quad q=|c|^2|\psi|^2+u_x+u^2
\]
solve equation
\[
 -2u_t=u_{xxx}+6(uu_{xx}+u^2_x+2u^2u_x)-3u_{xy}-6uu_y-6qu_x,\quad q_x=u_y.
\]
Of course this equation is linearizable as well: the Cole-Hopf
transformation $2u=v_x/v,$ $2q=v_y/v$ links it with the linear equation
$2v_t=3v_{xy}-v_{xxx}.$

%--------------------------------------------------------------------------
\paragraph{Acknowledgements.} I am grateful to S.V.~Manakov who drew my
attention to the paper \cite{Cheng-Li91} and B.G.~Konopelchenko,
A.B.~Shabat and R.I.~Yamilov for useful discussions and remarks. This work
was supported by grants RFBR-99-01-00431 and INTAS-99-01782.

%--------------------------------------------------------------------------

\end{document}